\documentclass[manuscript]{aastex}

\newcommand{\MO}{$M_\odot$}

\shorttitle{Line-of-sight Shell Structure of the Cygnus Loop}
\shortauthors{Uchida et al.}

\begin{document}


\title{Line-of-sight Shell Structure of the Cygnus Loop}


\author{Hiroyuki Uchida\altaffilmark{1}, Hiroshi Tsunemi\altaffilmark{1}, Satoru Katsuda\altaffilmark{1} \altaffilmark{2}, Masashi Kimura\altaffilmark{1}, Hiroko Kosugi\altaffilmark{1} and Hiroaki Takahashi\altaffilmark{1}}
\email{uchida@ess.sci.osaka-u.ac.jp}


\altaffiltext{1}{Department of Earth and Space Science, Graduate School of Science, Osaka University, Toyonaka, Osaka 560-0043, Japan}
\altaffiltext{2}{Code 662, NASA Goddard Space Flight Center, Greenbelt, MD 20771}


\begin{abstract}
We conducted a comprehensive study on the shell structure of the \object{Cygnus Loop} using 41 observation data obtained by the \textit{Suzaku} and the \textit{XMM-Newton} satellites.
To investigate the detailed plasma structure of the Cygnus Loop, we divided our fields of view into 1042 box regions.
From the spectral analysis, the spectra obtained from the limb of the Loop are well fitted by the single-component non-equilibrium ionization plasma model.
On the other hand, the spectra obtained from the inner regions are well fitted by the two-component model.
As a result, we confirmed that the low-temperature and the high-temperature components originated from the surrounding interstellar matter (ISM)  and the ejecta of the Loop, respectively.

From the best-fit results, we showed a flux distribution of the ISM component.
The distribution clearly shows the limb-brightening structure, and we found out some low-flux regions.
Among them, the south blowout region has the lowest flux.
We also found other large low-flux regions at slightly west and the northeast  from the center.
We estimated the former thin shell region to be $\sim1\arcdeg.3$ in diameter and concluded that there exists a blowout along the line of sight in addition to the south blowout. 
We also calculated the emission measure distribution of the ISM component and showed that the \object{Cygnus Loop} is far from the result obtained by a simple Sedov evolution model.
From the results, we support that the \object{Cygnus Loop} originated from a cavity explosion.
The emission measure distribution also suggests that the cavity-wall density is higher in the northeast than that in the southwest.
These results suggest that the thickness of the cavity wall surrounding the \object{Cygnus Loop} is not uniform.

\end{abstract}

\keywords{ISM: abundances --- ISM: individual (Cygnus Loop) ---
  supernova remnants --- X-rays: ISM}

\section{Introduction}\label{sec:intro}
The supernova (SN) explosions blow out the various heavy elements generated by the nucleosynthesis process inside the progenitor stars. Meanwhile, the blast wave generated by the SN explosion sweeps up and heats the interstellar matter (ISM). 
It forms a characteristic structure of each supernova remnant (SNR). 
In this way, the morphology of the shell structure provides information about the ambient density of the ISM.
 
The \object{Cygnus Loop} is one of the brightest SNRs in the X-ray sky. Its age is estimated to be $\sim10,000$yr and the distance is comparatively close to us (540pc; Blair et al. 2005). The large apparent size (2.5\arcdeg$\times$3.5\arcdeg; Levenson et al. 1997) enables us to study the plasma structure of the Loop. The origin of the \object{Cygnus Loop} is thought to be a cavity explosion \citep{McCray79, Hester86, Hester94, Levenson97}. The progenitor is presumed to be a B0, 15\MO \ star \citep{Levenson98} and some X-ray studies also estimated the progenitor mass to be 12-15\MO \ (\textit{e.g.}, Tsunemi et al. 2007).
From the morphological point of view, the \object{Cygnus Loop} is a typical shell-like SNR and it is almost circular in shape. 
However, a large break is seen in the south, known as ``blowout'' region \citep{Aschenbach99}.
The origin of the ``blowout'' is not well understood. \citet{Aschenbach99} explained this extended structure as a breakout into a lower density ISM. 
On the other hand, based on a radio observation, \citet{Uyaniker02} suggested the existence of a secondary SNR in the south.
Some other radio observations support this conclusion \citep{Uyaniker04, Sun06}. 
Recently, \citet{Uchida08} observed this region with the \textit{XMM-Newton} observatory and found that the X-ray spectra of this region consist of two plasma components with different electron temperatures. Judging from the plasma structures and the metal distributions, they concluded that the X-ray emission is consistent with a \object{Cygnus Loop} origin and that each plasma component is derived from the ejecta and the cavity material, respectively. They also showed that the X-ray shell is thin in their fields of view (FOV) and concluded that the origin of the blowout can be explained as a breakout into a lower density ISM as proposed by \citet{Aschenbach99}.

It is natural to consider that such a breakout may also exist along the line of sight. 
\citet{Tsunemi07} observed the \object{Cygnus Loop} along the diameter with \textit{XMM-Newton} and argued about it.
They divided their FOV into a north path and a south path and found the thin shell region to be 5\arcmin \ in the south path and 20\arcmin \ in the north path.
They estimated this thin shell region to have a diameter of 1\arcdeg, centering on $\alpha = 20^{\mathrm h}49^{\mathrm m}11^{\mathrm s}$, $\delta = 31\arcdeg05\arcmin20\arcsec$.
\citet{Kimura09} expanded their observation northward with \textit{Suzaku} and found that the flux of the swept-up matter in southwest is about a third of that in northeast. The width of this region is $\sim50\arcmin$.
\citet{Kimura09} presumed that there is a blowout region in the direction of our line of sight in the middle west of the Loop.

In this paper, we used 41 observation data obtained by the \textit{Suzaku} and the \textit{XMM-Newton} observatories. We reanalyzed all the data to conduct a comprehensive study on the shell structure of the \object{Cygnus Loop}.

\section{Observations}
\begin{figure}
  \begin{center}
    \includegraphics[width=120mm]{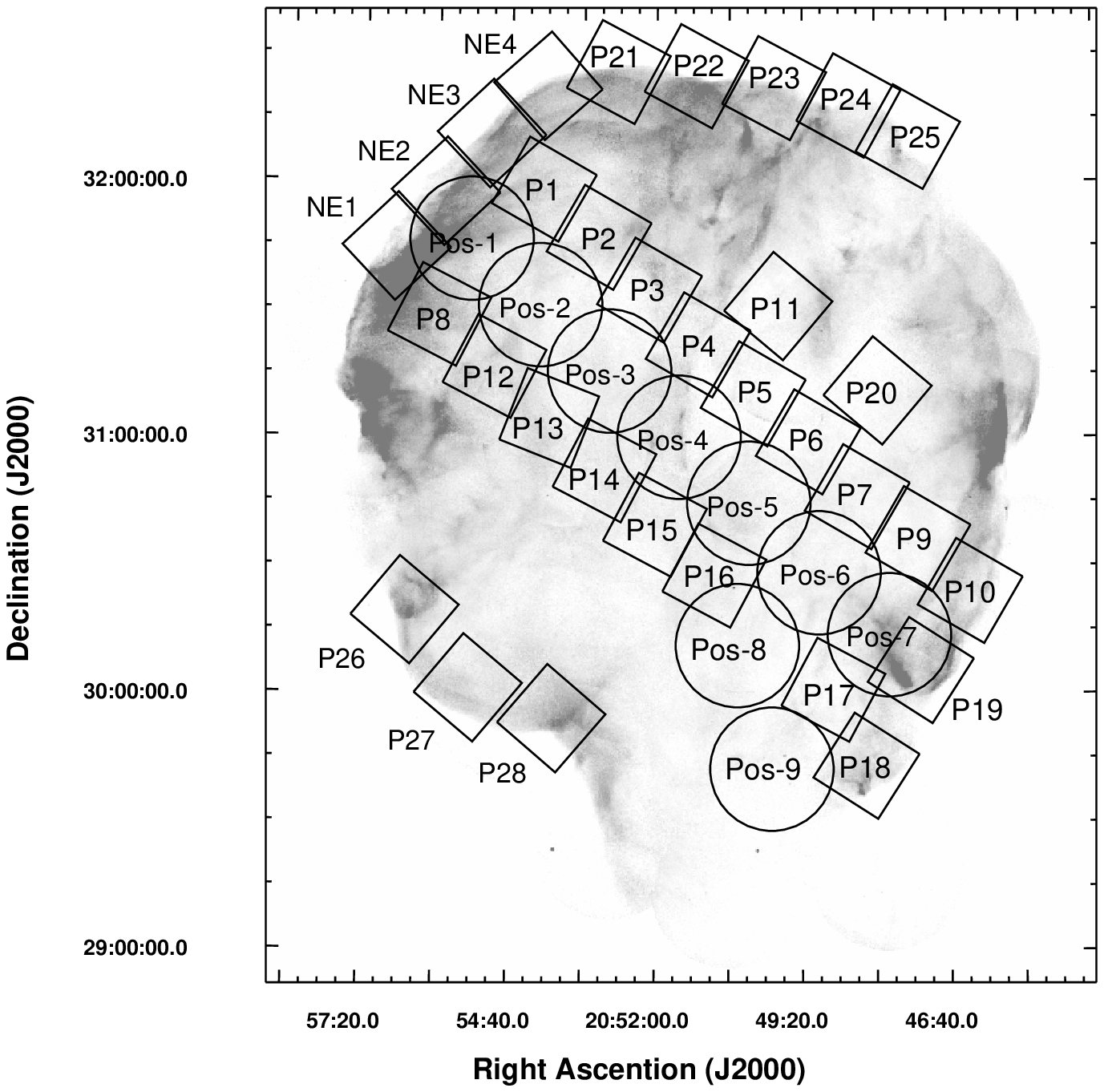}
    \includegraphics[width=120mm]{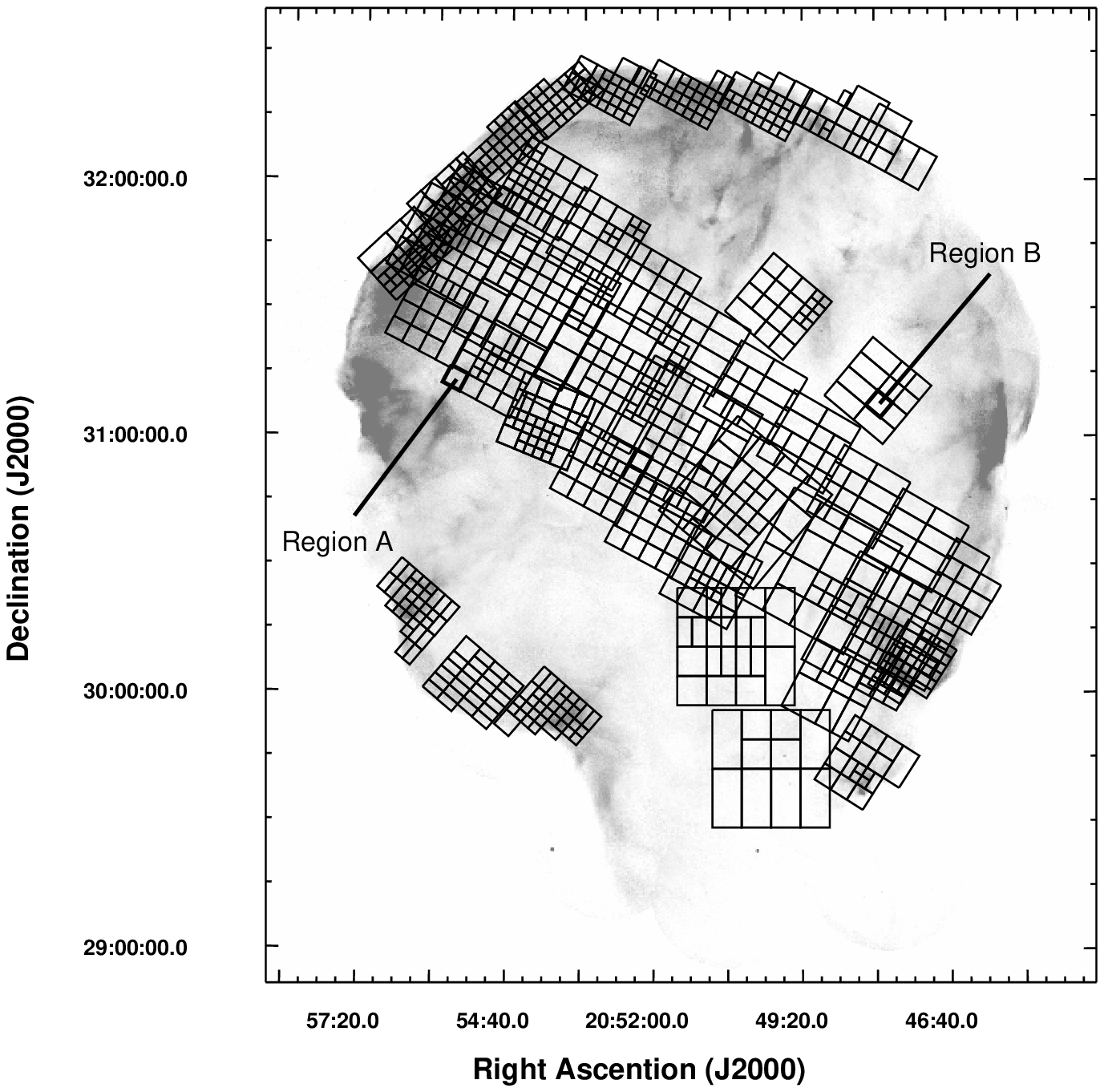}
  \end{center}
  \caption{\textit{Top}: \textit{ROSAT} HRI image of the entire Cygnus Loop. The circles and rectangles represent our FOV of the \textit{XMM-Newton} MOS  and the \textit{Suzaku} XIS, respectively. \textit{Bottom}: Same as the left panel, but for overlaid with the spectral extraction regions with  small rectangles.}\label{fig:HRI}
\end{figure}

We summarized the 41 observations in table \ref{tab:sum}. 
All the data were taken between 2002 and 2008. 
The observing regions are shown in figure \ref{fig:HRI} top. 
The circles and rectangles represent the FOV of the \textit{XMM-Newton} MOS  and the \textit{Suzaku} XIS, respectively.

All of the \textit{Suzaku} data were analyzed with version 6.5 of the HEAsoft tools. 
For reduction of the \textit{Suzaku} data, we used version 9 of the Suzaku Software. 
The calibration database (CALDB) used was updated in July 2008. 
We used revision 2.2 of the cleaned event data and combined 3$\times$3 and 5$\times$5 event files. 
We applied the spaced row charge injection (SCI) method \citep{Prigozhin08} to P1, P2, P3, P4, P5, P6, P7, P9, P10, P11, P18, P19, P20, P21, P22, P23, P24 and P25 data. 
This method reduces the effect of radiation damage of the XIS and recovers the energy resolution, for example, from 205$\pm6$eV to 157$\pm4$eV at 5.9 keV. 
In order to exclude the background flare events, we obtained good time intervals (GTIs) including only times at which the count rates were within $\pm2\sigma$ of the mean count rates.

Since the \object{Cygnus Loop} is a large diffuse source and our FOV are filled with the SNR's emission, we could not  obtain background spectra from our FOV. We also had no background data from the neighborhood of the \object{Cygnus Loop}. We therefore applied the Lockman Hole data for background subtraction. We reviewed the effect of the galactic ridge X-ray emission (GRXE). The flux of the GRXE at $l = 62\arcdeg$, $|b| < 0\arcdeg.4$ is $6\times10^{-12}$erg cm$^{-2}$s$^{-1}$deg$^{-2}$ (0.7-2.0 keV) \citep{Sugizaki01}. Although the \object{Cygnus Loop} ($l = 74\arcdeg$, $b = -8\arcdeg.6$) is located outside the FOV of \citet{Sugizaki01}, this value gives us an upper limit of the GRXE at the \object{Cygnus Loop}. 
Meanwhile, the flux of the \object{Cygnus Loop} is estimated to be $8.2\times10^{-10}$erg cm$^{-2}$s$^{-1}$deg$^{-2}$ (0.7-2.0 keV), assuming that the \object{Cygnus Loop} is a circle with a diameter of $3\arcdeg.0$. 
Therefore, we concluded that the effect of the GRXE on the \object{Cygnus Loop} is vanishingly small. 
The solar wind charge exchange (SWCX) is also considered to correlate with the soft X-ray background below 1 keV  \citep{Fujimoto07}. However, in terms of the \object{Cygnus Loop}, we consider that the SWCX is negligible because of the prominent surface brightness of the Loop. Thus, the Lockman Hole data obtained in 2006, 2007 and 2008 were applied for background subtraction. 
We selected data whose observation dates were close to those of the \object{Cygnus Loop} observations. Since there were no photons above 3.0 keV after background subtraction, the energy ranges of 0.2-3.0 keV and 0.4-3.0 keV were used for XIS1 (back-illuminated CCD; BI CCD) and XIS0,2,3 (front-illuminated CCD; FI CCD), respectively \citep{Koyama07}. 

All the \textit{XMM-Newton} data were processed with version 7.1.0 of the \textit{XMM} Science Analysis System (SAS). The current calibration files (CCFs) used were updated on 2008 June. We used data obtained with the EPIC MOS and pn cameras. These data were taken by using the medium filters and the prime full-window mode. We selected X-ray events corresponding to patterns 0-12 and flag = 0 for MOS 1 and 2, patterns 0-4 and flag = 0 for pn, respectively. In order to exclude background flare events, we determined the GTIs in the same way as those of the \textit{Suzaku} data. After filtering the data, they were vignetting-corrected by using the SAS task \textbf{evigweight}. For background subtraction, we employed blank-sky observations prepared by \citet{Read03} for a similar reason with the \textit{Suzaku} case. After the background subtraction, the energy range of 0.3-3.0 keV was used for each instrument.

\section{Spectral Analysis}\label{sec:specana}
To investigate the plasma structure of the \object{Cygnus Loop}, we divided the entire FOV into small box regions. In order to equalize the statistics, we initially divided all images of XIS1 or MOS2 into two parts; if each divided region had more than 10,000 photons, it was once again divided. In this way, we obtained 1042 box regions. Each region contained 5,000-10,000 photons for XIS1 and MOS2. The side length of each box ranges from 2.2\arcmin \ to 14\arcmin. Therefore box sizes are not smaller than the angular resolution capability of the \textit{Suzaku} XIS. We grouped 1042 spectra into bins with a minimum of 20 counts such that $\chi^2$ statistics are appropriate. 
 In order to generate a response matrix file (RMF) and an ancillary response file (ARF), we employed xisrmfgen \citep{Ishisaki07} and xissimarfgen for the \textit{Suzaku} data, rmfgen and arfgen for the XMM-Newton data.

Firstly, we fitted all the spectra by a single-component variable abundance non-equilibrium ionization (VNEI) model. 
We employed \textbf{TBabs} (T\"{u}bingen-Boulder ISM absorption model; Wilms et al. 2000) and \textbf{VNEI} (NEI ver.2.0; Borkowski et al. 2001) in XSPEC version 12.5.0 \citep{Arnaud96}.
In this model, the abundances for C, N, O, Ne, Mg, Si and Fe were free, while we set the relative abundances for S to the solar value equal to that of Si, Ni equal to Fe. 
The other elements were fixed to their solar values.
Other parameters were all free such as the electron temperature, $kT_e$, the ionization timescale, $\tau$ (a product of the electron density and the elapsed time after the shock heating), and the emission measure, EM ($=\int n_e n_{\rm H} dl$, where $n_e$ and $n_{\rm H}$ are the number densities of electron and hydrogen and $dl$ is the plasma depth). 
We also set the absorption column density, $\rm\textit{N}_H$, to be free.
As a result, the spectra from the limb regions are well fitted by the single-component VNEI model.
As shown by earlier observations of the northeast and the southeast limb \citep{Tsunemi07, Kimura09, Uchida09Nrim, Tsunemi09}, the spectra obtained from the limb regions of the Cygnus Loop are typically described by a single-component VNEI model.

On the other hand, the spectra from the inner regions are generally not fitted by the single-component VNEI model.
From earlier observations of the northeast to the southwest regions along the diameter, \citet{Tsunemi07} found that the spectra from the inner regions of the \object{Cygnus Loop} consist of a two-component VNEI plasma.
They concluded the plasma structure of the \object{Cygnus Loop} as follows: the high-$kT_e$ ejecta component is surrounded by a low-$kT_e$ ISM component. \cite{Uchida09ejecta} showed that the two-component VNEI model is wholly applicable to the inner regions of the \object{Cygnus Loop}. 
Therefore, we next intended to give an additional high-$kT_e$ VNEI component to the single-component VNEI model.
In this model, we fixed the metal abundances of the low-$kT_e$ component to the values obtained from the result of \citet{Tsunemi07}, since the model whose abundances set all free could not obtain the physically meaningful results. 
\citet{Tsunemi07} showed the relative abundances to the solar values of the ISM component as follows: C=0.27, N=0.10, O=0.11, Ne=0.21, Mg=0.17, Si=0.34, S=0.17, Fe(=Ni)=0.20. In addition, they fixed other elements to the solar values \citep{Anders89}. 
Meanwhile, in the high-$kT_e$ component, the abundances for O, Ne, Mg, Si, and Fe were free, while we set the abundances for C and N equal to O, S equal to Si, Ni equal to Fe. Other elements were fixed to their solar values.
The other parameters such as  $kT_e$, $\tau$, EM, and $\rm\textit{N}_H$ were all free.
We applied both single-component VNEI model and two-component VNEI model to all the spectra and determined which model is acceptable by using the F-test with a significance level of 99\%.
As a result, roughly $<0.80R_{\rm s}$ of the northeast region and $<0.85R_{\rm s}$ of the southwest region need an additional component, where $R_{\rm s}$ is a shock radius.
Here, we define the ``limb observations'' as the regions where the single-component VNEI model is acceptable and the ``inside observations'' as the remaining regions.

Figure \ref{fig:spec} shows two example XIS1 spectra.
The spectral extracted regions are shown in figure \ref{fig:HRI}.
Both regions are located at the inside observations and the two-component VNEI model is applicable.
The bottom two panels show the best-fit results of the two-component VNEI model.
Blue and red lines represent the high-$kT_e$ and the low-$kT_e$ component. 
We also show the result with the single-component VNEI model at the top two panels for comparison.
The best-fit parameters are shown in table \ref{tab:spec}.
These results show that  the reduced $\chi^2$ values are significantly improved with the two-component VNEI models.

\section{Discussion}
\subsection{Temperature distribution of the low-$kT_e$ component}
All the spectra are well fitted by either the single-component VNEI model or the two-component VNEI model.
From the best-fit parameters of the inside observations, we found that the electron temperature of the low-$kT_e$ component is almost uniform.
The averaged value is 0.23 keV ($\sigma = 0.08$ keV) and it is sufficiently lower than that of the high-$kT_e$ component (0.52 keV, $\sigma = 0.17$ keV).
The temperature of the low-$kT_e$ component is close to that of the limb observations (0.29 keV, $\sigma = 0.07$ keV).
Therefore we collectively call these components ``low-$kT_e$ component'' hereafter.

\begin{figure}
  \begin{center}
     \includegraphics[width=120mm]{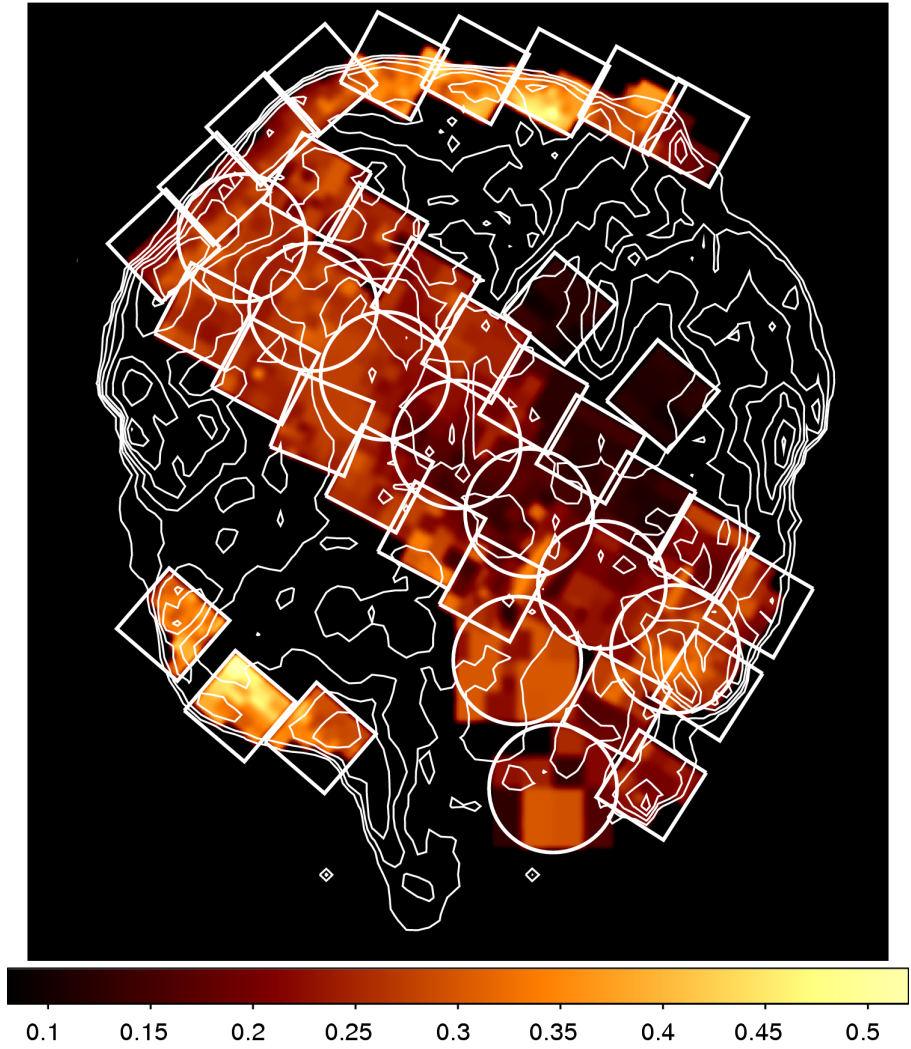}
 \end{center}
  \caption{Our FOV and the electron temperature distribution of the low-$kT_e$ component overlaid with the white contour from the \textit{ROSAT} HRI image. The images are smoothed by Gaussian kernel of $\sigma=2.8\arcmin$. The values are in units of keV.}\label{fig:kTe}
\end{figure}

Figure \ref{fig:kTe} shows our FOV and the electron temperature distribution of the low-$kT_e$ component overlaid with the white contour from the \textit{ROSAT} HRI image.
The averaged value is $\sim0.28$ keV and it ranges from 0.12 keV to 0.35 keV. 
Meanwhile, the temperature of the high-$kT_e$ component ranges from 0.4 keV to 0.9 keV, which is consistent with the previous observations \citep{Tsunemi07, Katsuda08ejecta, Kimura09, Uchida09ejecta}.
Then, we confirmed that  the temperature of each component is clearly separated.
\citet{Uchida09ejecta} also showed that the temperature distribution of the high-$kT_e$ component is not uniform and that it is lower in the southwest part than that in the northeast part.
On the other hand,  the temperature of the low-$kT_e$ component is relatively uniform (see figure \ref{fig:kTe}).
The detailed distribution shows the temperature near the center is lower than that of the surroundings.
We also found that the temperature distribution is seamless at the boundary between the limb observations and the inside observations.
Therefore, the low-$kT_e$ components of these regions must have the same origin.
The spectra from the limb observations are obviously swept-up ISM origin, and thus, we concluded that any low-$kT_e$ component originates from the ISM component.

\subsection{Line-of-sight Shell Structure of the Cygnus Loop}
\begin{figure}
  \begin{center}
    \includegraphics[width=80mm]{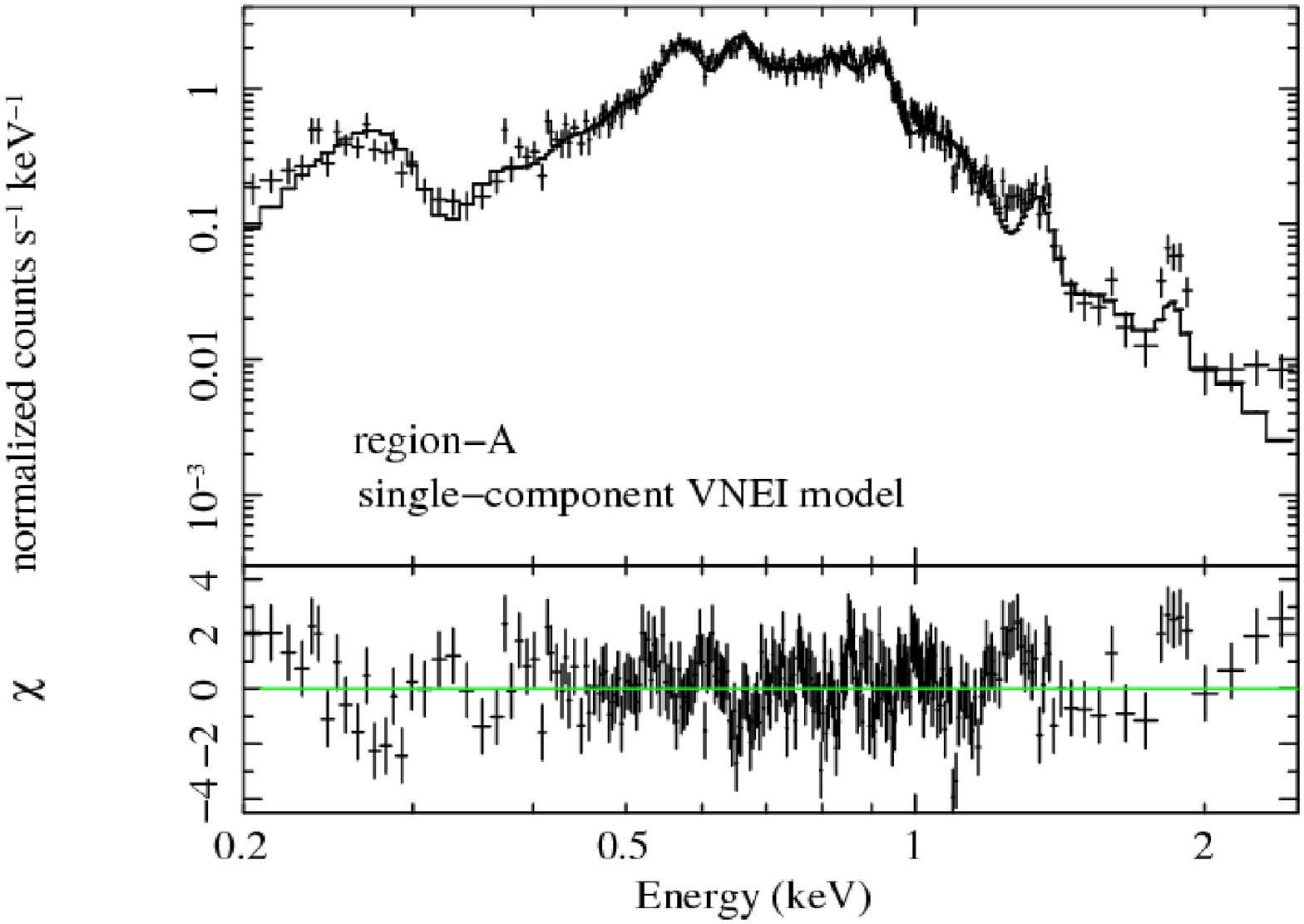}
    \includegraphics[width=80mm]{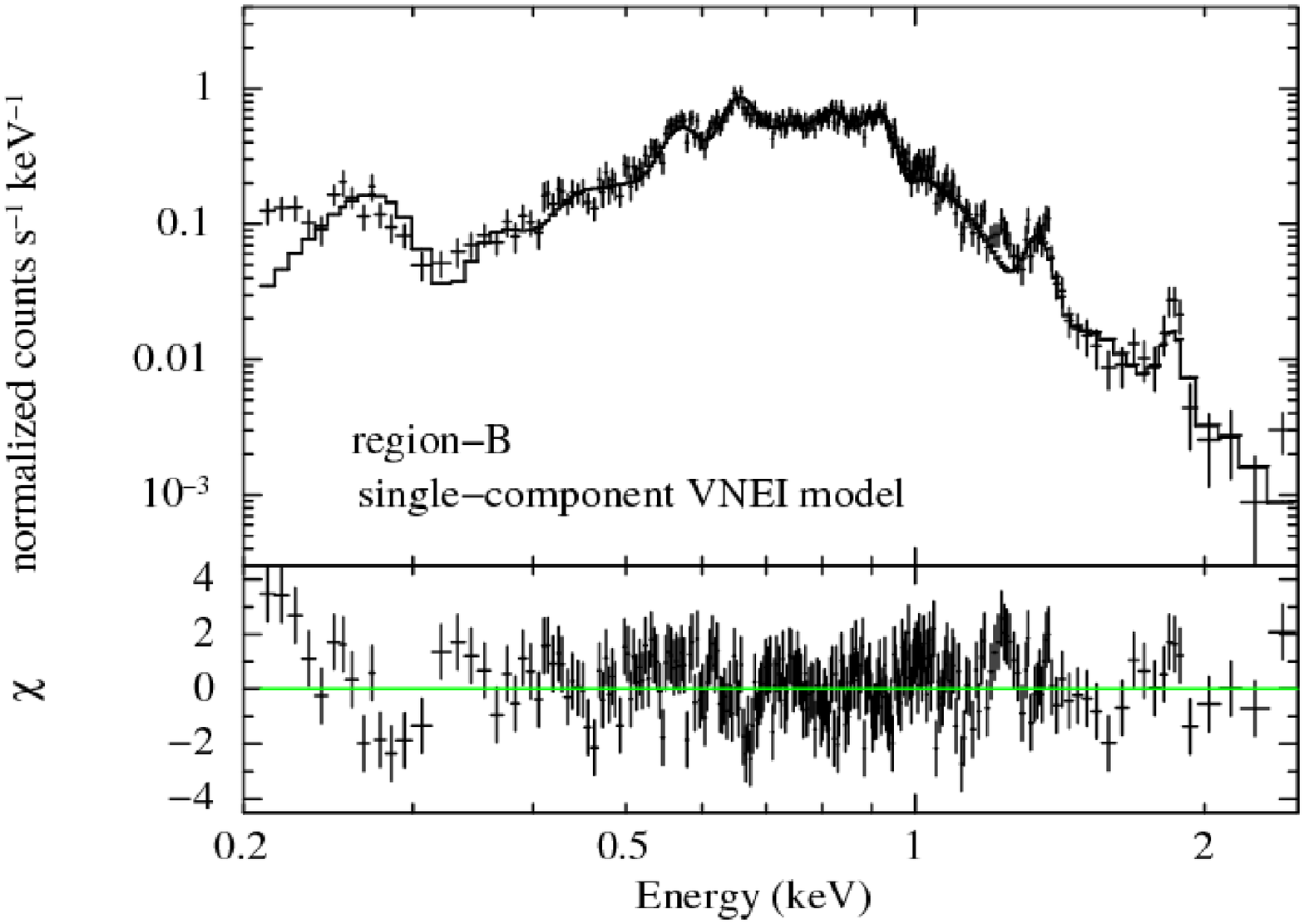}
    \includegraphics[width=80mm]{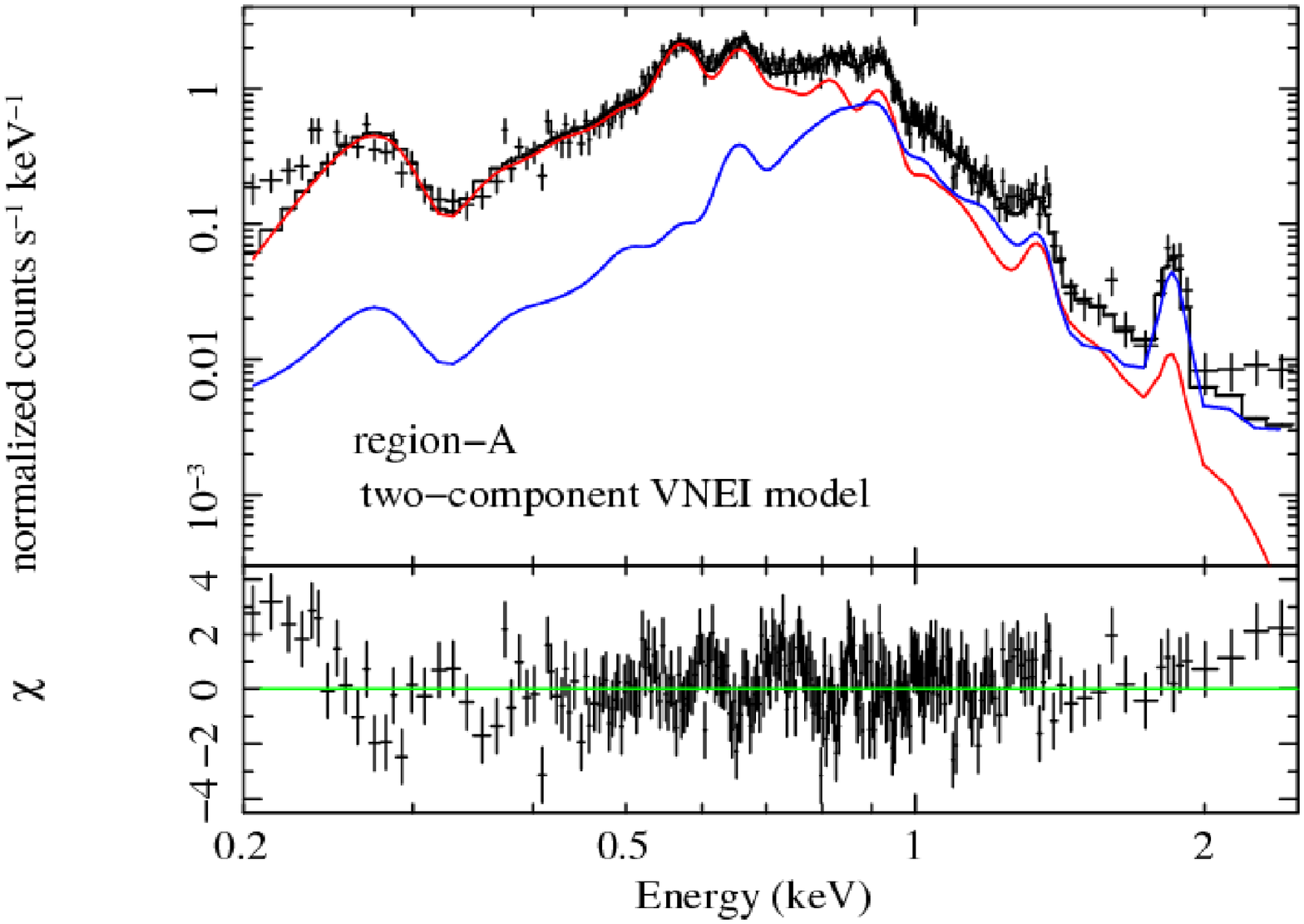}
    \includegraphics[width=80mm]{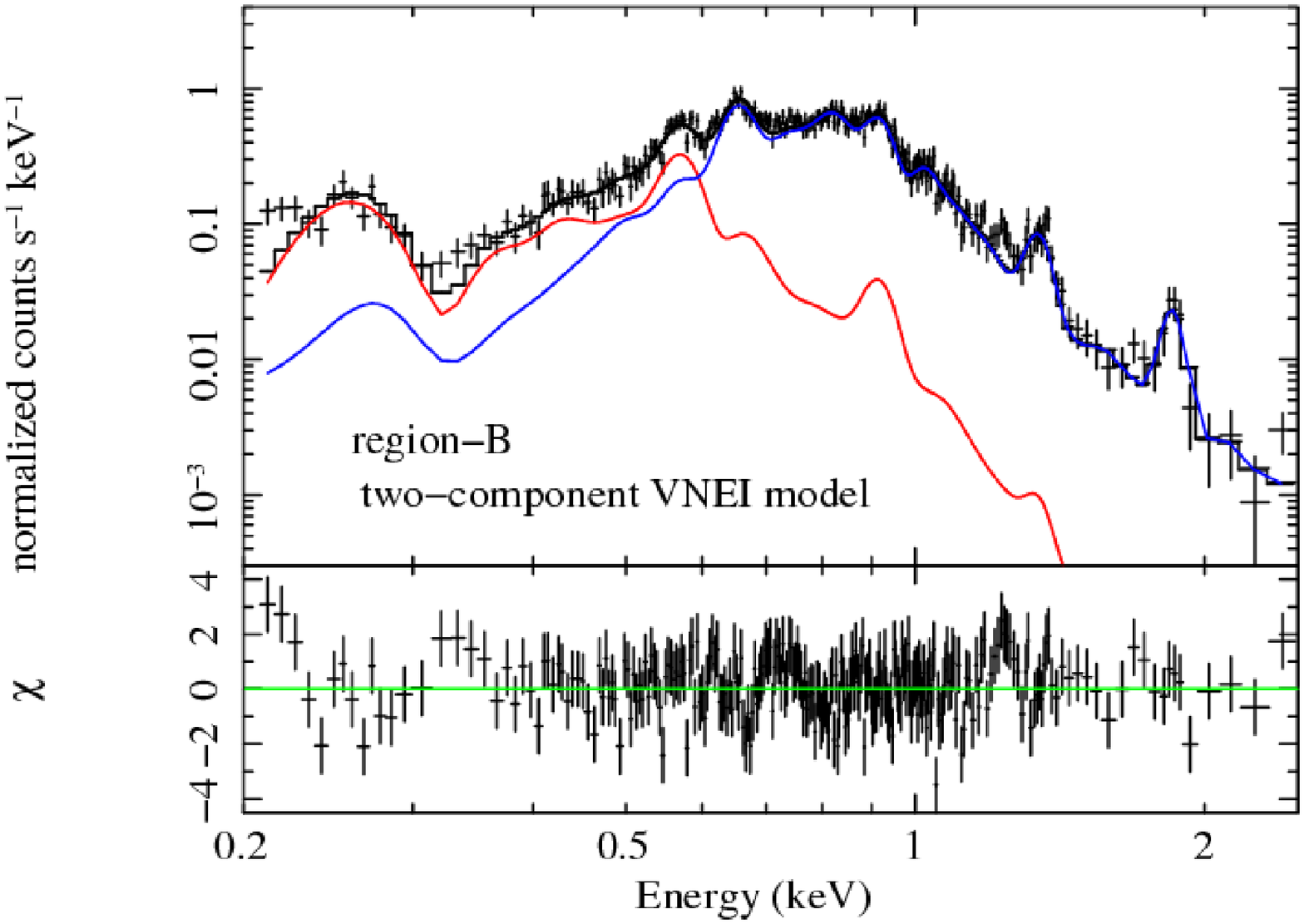}
  \end{center}
  \caption{Example XIS1 spectra from the regions where the flux of the swept-up matter is high (region-A: left two panels) and low (region-B: right two panels), respectively (see figure\ref{fig:HRI}). The best-fit curves for the single-component VNEI models are shown by solid black lines in the top two panels.  Bottom two panels are the same as the top panels, but for the fitting results with the two-component VNEI models. In the bottom panels, blue and red lines represent the high-$kT_e$ component and the low-$kT_e$ component, respectively. The residuals are shown in lower panels.}\label{fig:spec}
\end{figure}

Taking into account the age of the \object{Cygnus Loop} the reverse shocks should have already reached its center.
Therefore, on the assumption that the density of the ejecta-origin plasma is homogeneous, the X-ray flux depends exclusively on its plasma depth.
In figure \ref{fig:spec}, the blue line represents the high-$kT_e$ component of the two-component VNEI model.
Since the region-A and the region-B are located at the same radial distance from the center ($R\sim50\arcmin$, where we define $R$ as a distance from the ``geometric center'' determined by Levenson et al. 1998), they should have almost the same plasma depths.
Accordingly, the fluxes of the high-$kT_e$ components are actually not so different, while the spectral extracted regions are separated.
Meanwhile, the contributions of the low-$kT_e$ components are quite different as shown with the red lines in figure \ref{fig:spec}.
From the bottom left panel of figure \ref{fig:spec}, the flux of the low-$kT_e$ component in the region-A overwhelms that of the high-$kT_e$ component at 0.2-1.0 keV.
On the other hand, the contribution of the low-$kT_e$ component in the region-B is clearly smaller than that in the region-A.
Such a difference should be attributed to the difference of the surrounding shell of each region.
The value of the flux is proportional to EM ($\propto n_{\rm H}^2l$), which means that the surface brightness is sensitive to the change of the density and the plasma depth there.

\begin{figure}
  \begin{center}
    \includegraphics[width=160mm]{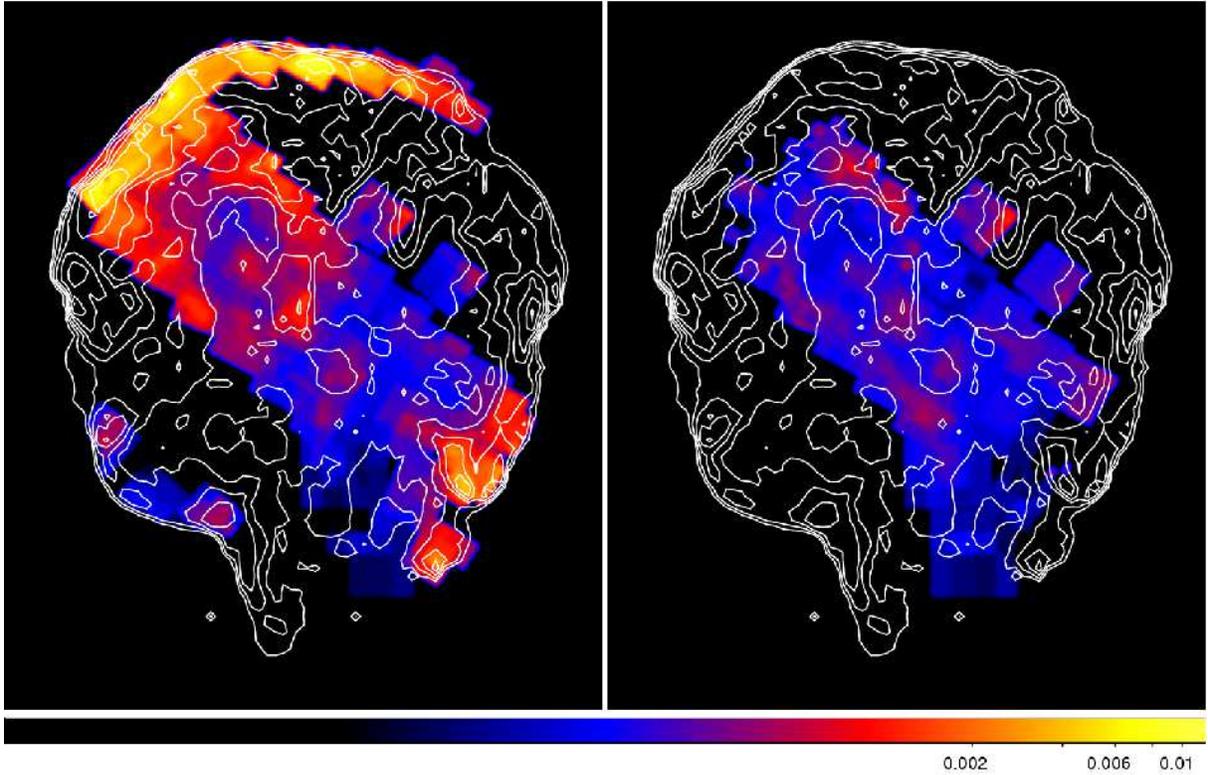}
  \end{center}
  \caption{0.2-3.0  keV flux distribution of the low-$kT_e$ (left) and the high-$kT_e$ (right) component in logarithmic scales overlaid with the white contour of the \textit{ROSAT} HRI image. The images are smoothed by Gaussian kernel of $\sigma=2.8\arcmin$. The values are in units of counts cm$^{-2}$s$^{-1}$arcmin$^{-2}$ and the scale parameters correspond with each other. Blue and red correspond to $\sim10^{-4}$ and $\sim10^{-3}$ counts cm$^{-2}$s$^{-1}$arcmin$^{-2}$, respectively.}\label{fig:flux}
\end{figure}

In order to estimate the ambient density of the \object{Cygnus Loop}, we calculated the fluxes of the low-$kT_e$ component from all regions.
Figure \ref{fig:flux} left shows the 0.2-3.0 keV flux distribution of the low-$kT_e$ component.
We also show that of the high-$kT_e$ component at the right panel.
The flux distribution of the high-$kT_e$ component is relatively uniform compared with that of the low-$kT_e$ component.
It reflects that the ejecta component uniformly filled inside the Loop.
In contrast, from the left panel, we clearly see the "limb-brightening" which reflects the spherical shell structure. 
Therefore, we confirmed that the low-$kT_e$ component comes from the surrounding ISM.
We also found that the northeast flux is higher than that in the southwest. 
It suggests that the density is higher in the direction of the northeast than that of the southwest.
The detailed shell structures are also seen from the left panel, for example, the ``V-shape'' knot at the southwest \citep{Aschenbach99, Leahy04}.

From the left panel of figure \ref{fig:flux}, we found the flux distribution inside the Loop is far from what we expect in the uniform shell structure.
This suggests the ambient density and the shell thickness varies considerably from region to region.
Thus, we can study the line-of-sight shell structure of the Loop.
Considering the relation between the surface brightness and the plasma density, the flux of the low-$kT_e$ component reflects the local density of the ISM. 
For example, the bright region in the northeast part is considered that the blast waves are expanding into the dense ISM there.
In contrast, there is a low-flux region at the south of the Loop (see figure \ref{fig:flux}).
It suggests the ambient density there is extremely low compared with other areas of the Loop.
As shown by \citet{Uchida08}, we noticed that there is a large break in the south where the ISM density is very thin.
In general, the velocity of the blast wave toward such tenuous ISM should become higher than other region.
Therefore, it forms a blowout where the shell thickness must be thin.

From figure \ref{fig:flux}, we also found a large low-flux region at slightly west of the \object{Cygnus Loop} center.
Although our FOV does not cover the whole region, the structure is close to a circular form, and we estimated the diameter to be $\sim1.3\arcdeg$.
The size is comparable to that of the south blowout.
The existence of such large low-flux region suggests that it has a blowout structure along the line of sight like the south blowout.
This result confirms the prediction by \citet{Tsunemi07} and \citet{Kimura09}. 
From figure \ref{fig:flux}, the northeast of the center also has lower flux than that of the surrounding region.
It strongly indicates that the line-of-sight ambient density there is locally low as well as that in the south blowout. 
This region has a C-shape structure which could be explained by the superposition of the circular low-flux region and the bright region where the blast wave interacts with a small cloud.
We estimate the diameter of this low-flux region to be  $\sim30\arcmin$.
These results show the ambient density of the \object{Cygnus Loop} is quite different from region to region.

\subsection{Evidence of Cavity Explosion}
\begin{figure}
  \begin{center}
    \includegraphics[width=150mm]{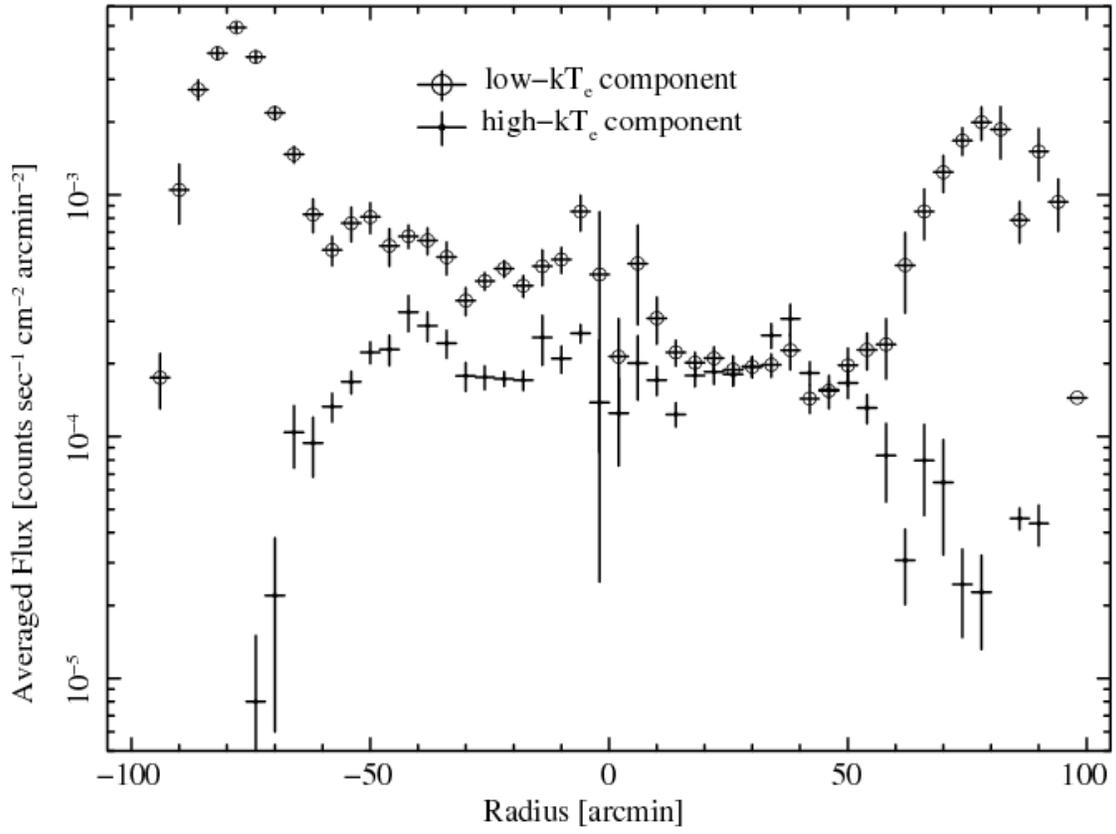}
  \end{center}
  \caption{Averaged flux profile as a function of $R$. The circles and triangles represent the flux of low-$kT_e$ and high-$kT_e$ components, respectively.}\label{fig:flux_plot}
\end{figure}

To put our result into perspective, we plotted the flux of each component as a function of radius $R$ as shown in figure \ref{fig:flux_plot}.
From this figure, we found the flux of the high-$kT_e$ component (shown as crosses) decreases from the center to the outside, which reflects the spherical structure of the ejecta filled inside the \object{Cygnus Loop}.
On the other hand, the flux of the low-$kT_e$ component (circles) has a limb-brightening structure, as mentioned in the previous section.
Furthermore the low-$kT_e$ flux at the southwest ($R>0$) is totally lower than that at the northeast.
While the high-$kT_e$ flux distribution is approximately symmetric, the low-$kT_e$ flux  is a few times higher at the northeast than that at the southwest.
In addition, looking at the inner region of the Loop, the flux distribution of the low-$kT_e$ component is declining from $R=-50$ to $R=50$.
This fact suggests the ambient density of the \object{Cygnus Loop} globally decreases from the northeast to the southwest.

\begin{figure}
  \begin{center}
    \includegraphics[width=160mm]{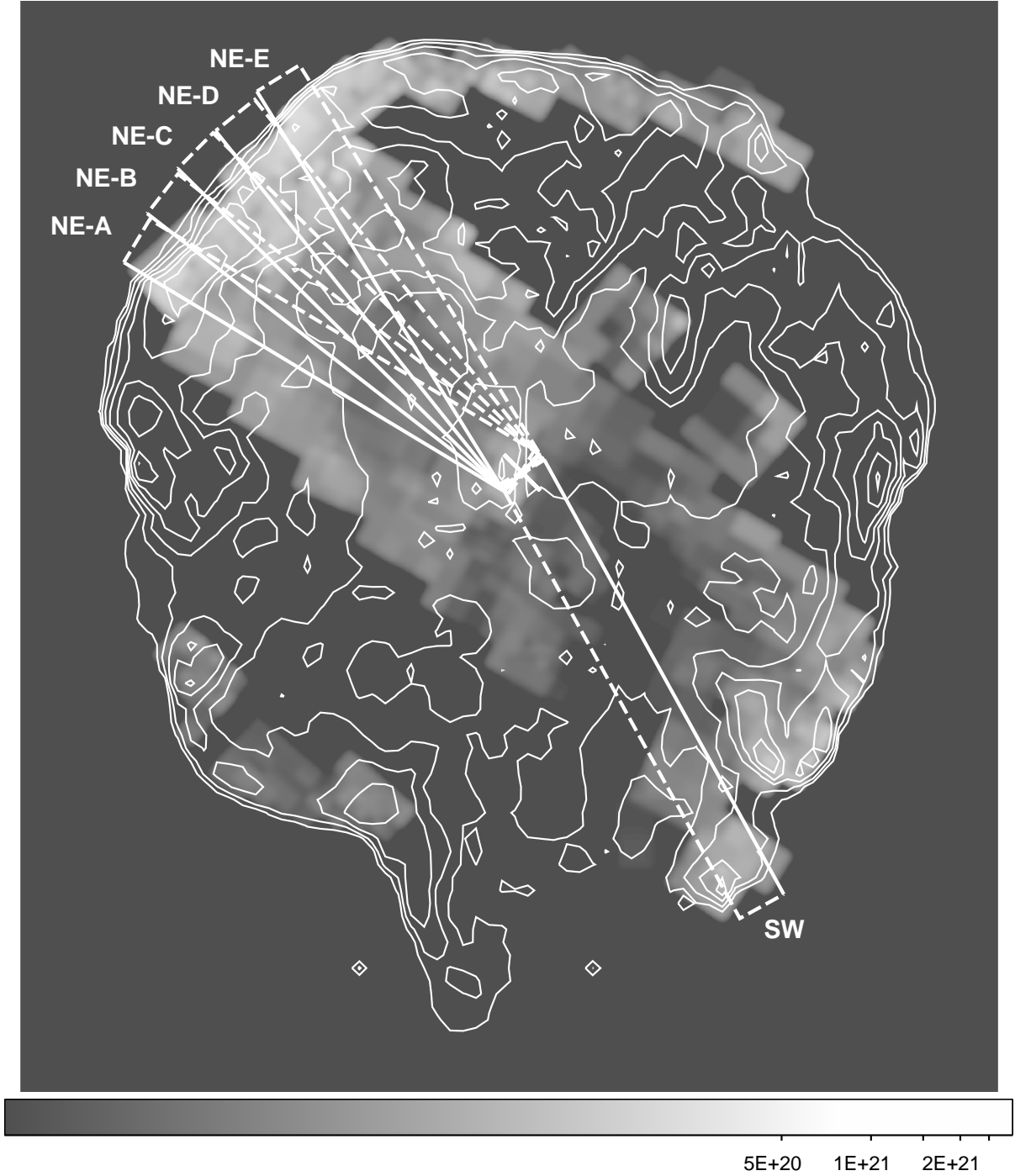}
  \end{center}
  \caption{EM  distribution of the low-$kT_e$ component in logarithmic scales overlaid with the white contour of the \textit{ROSAT} HRI image.}\label{fig:EM_region}
\end{figure}

\begin{figure}
  \begin{center}
    \includegraphics[width=75mm]{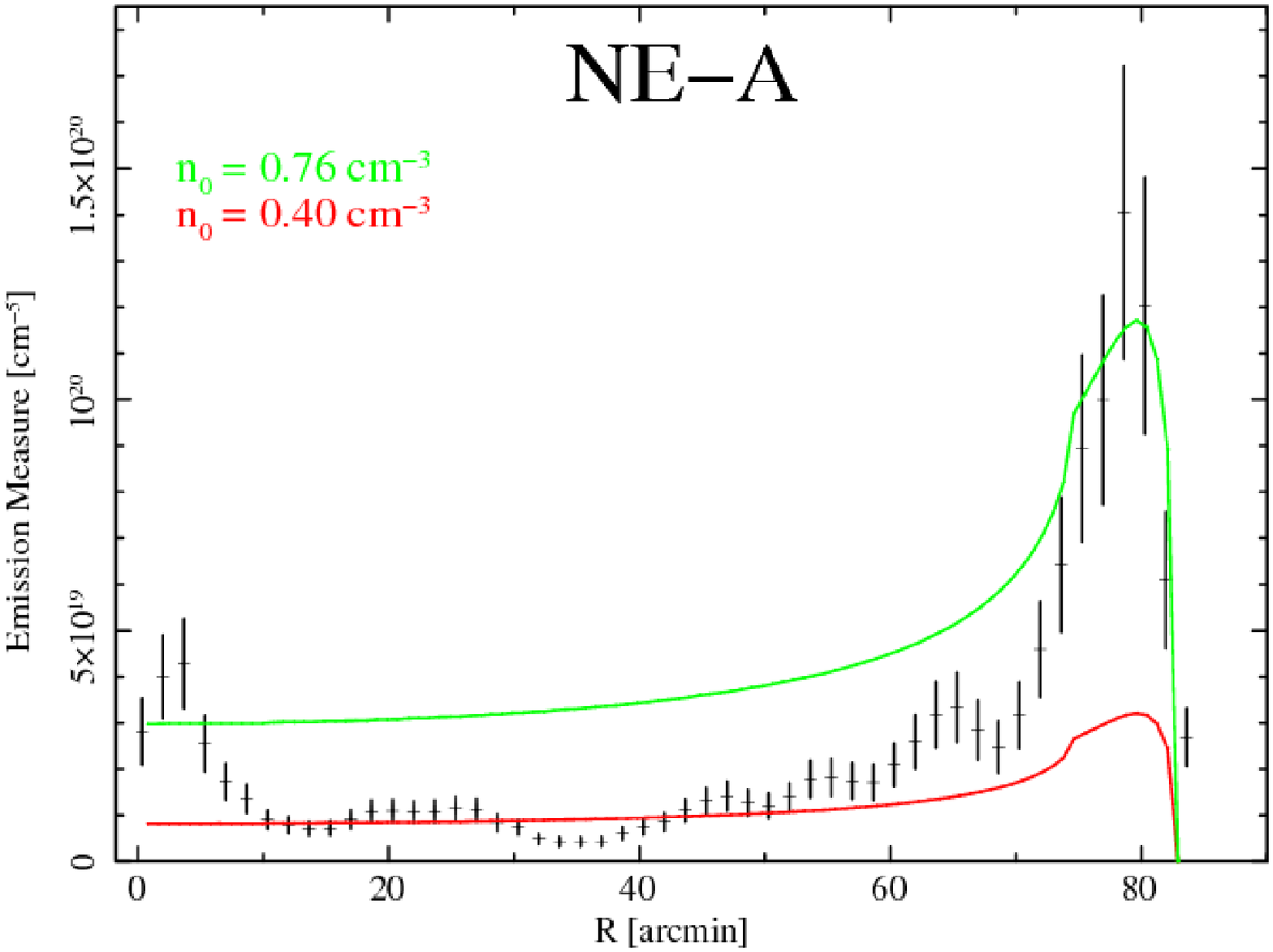}
    \includegraphics[width=75mm]{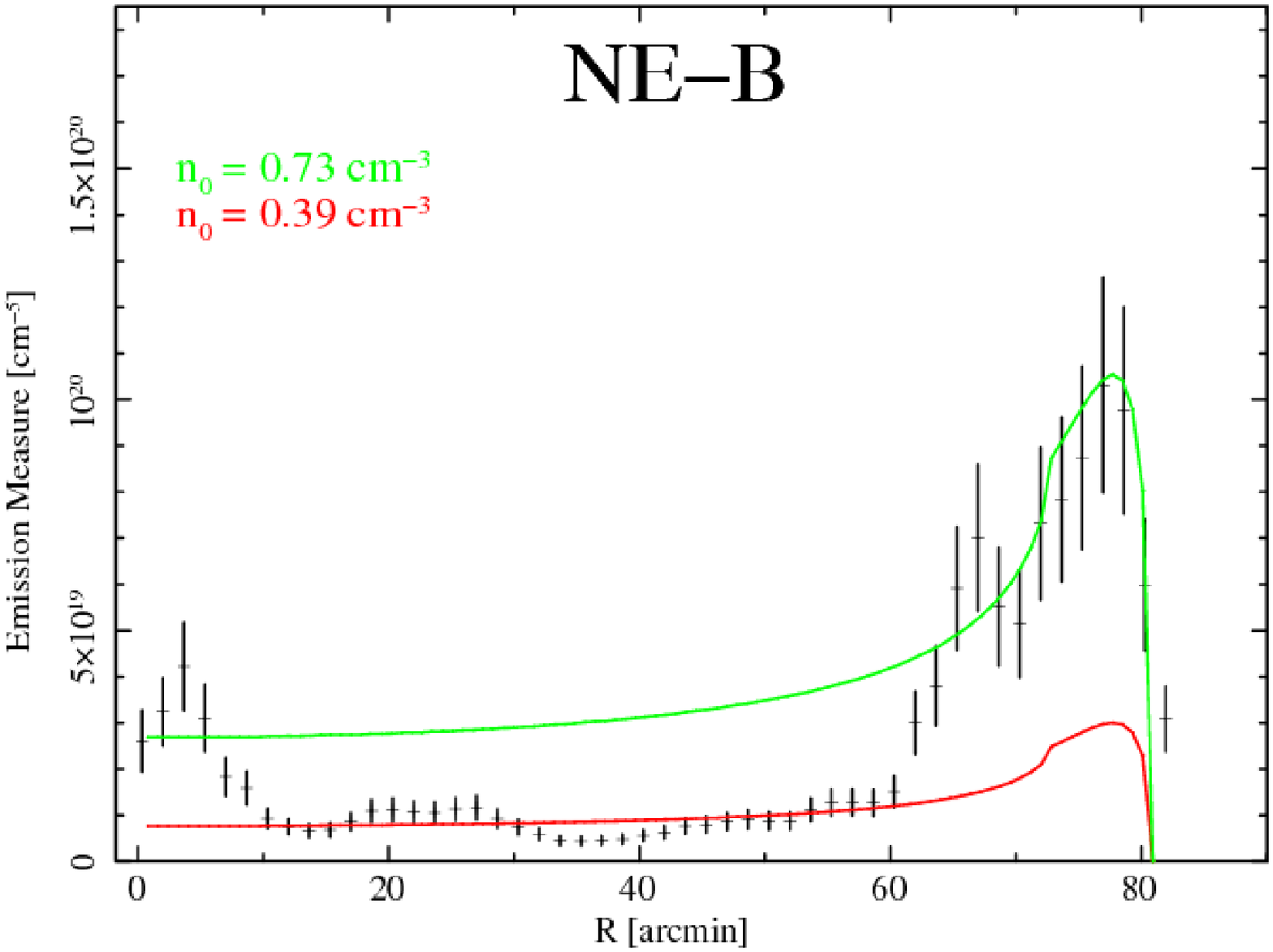}
    \includegraphics[width=75mm]{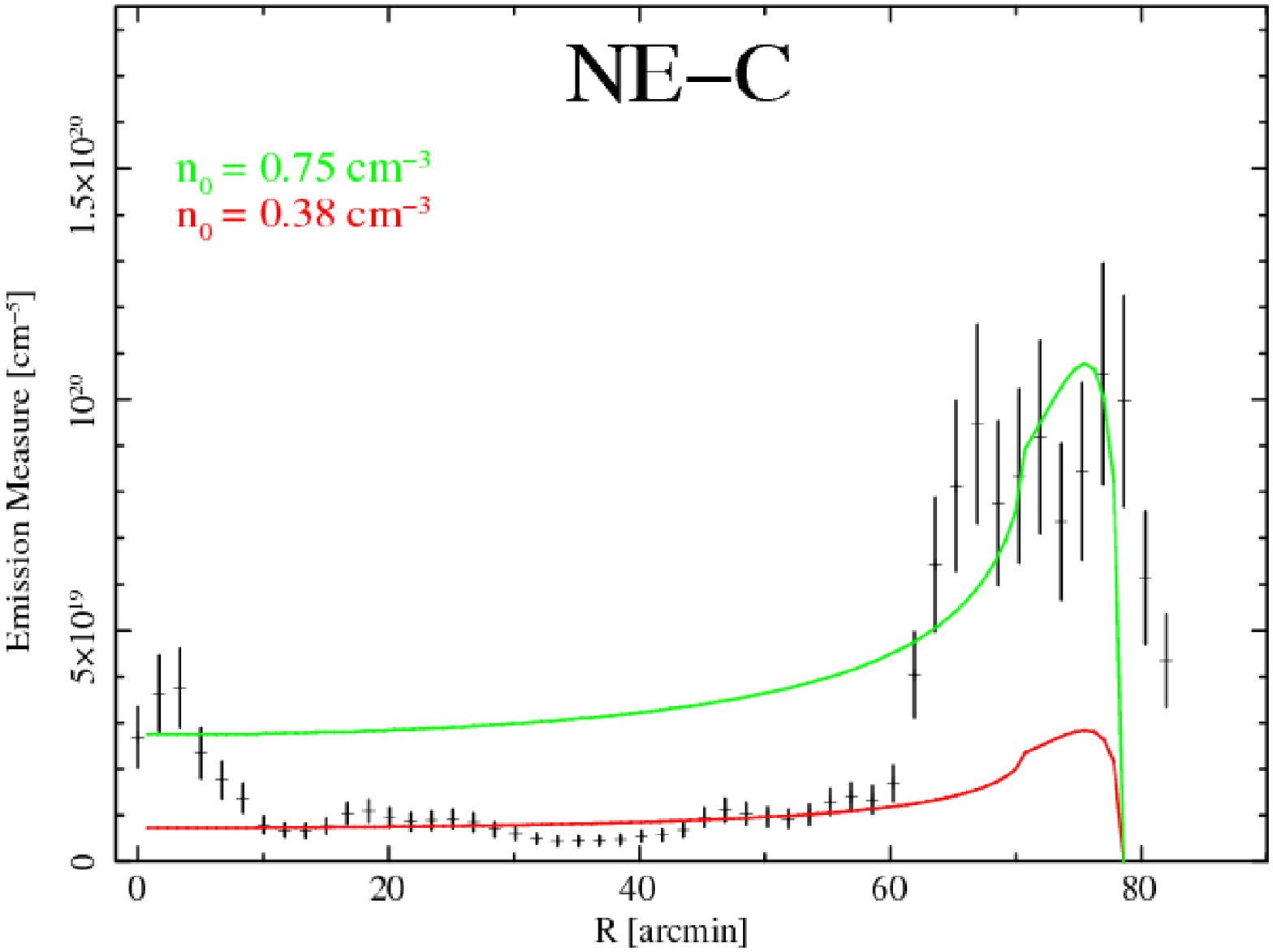}
    \includegraphics[width=75mm]{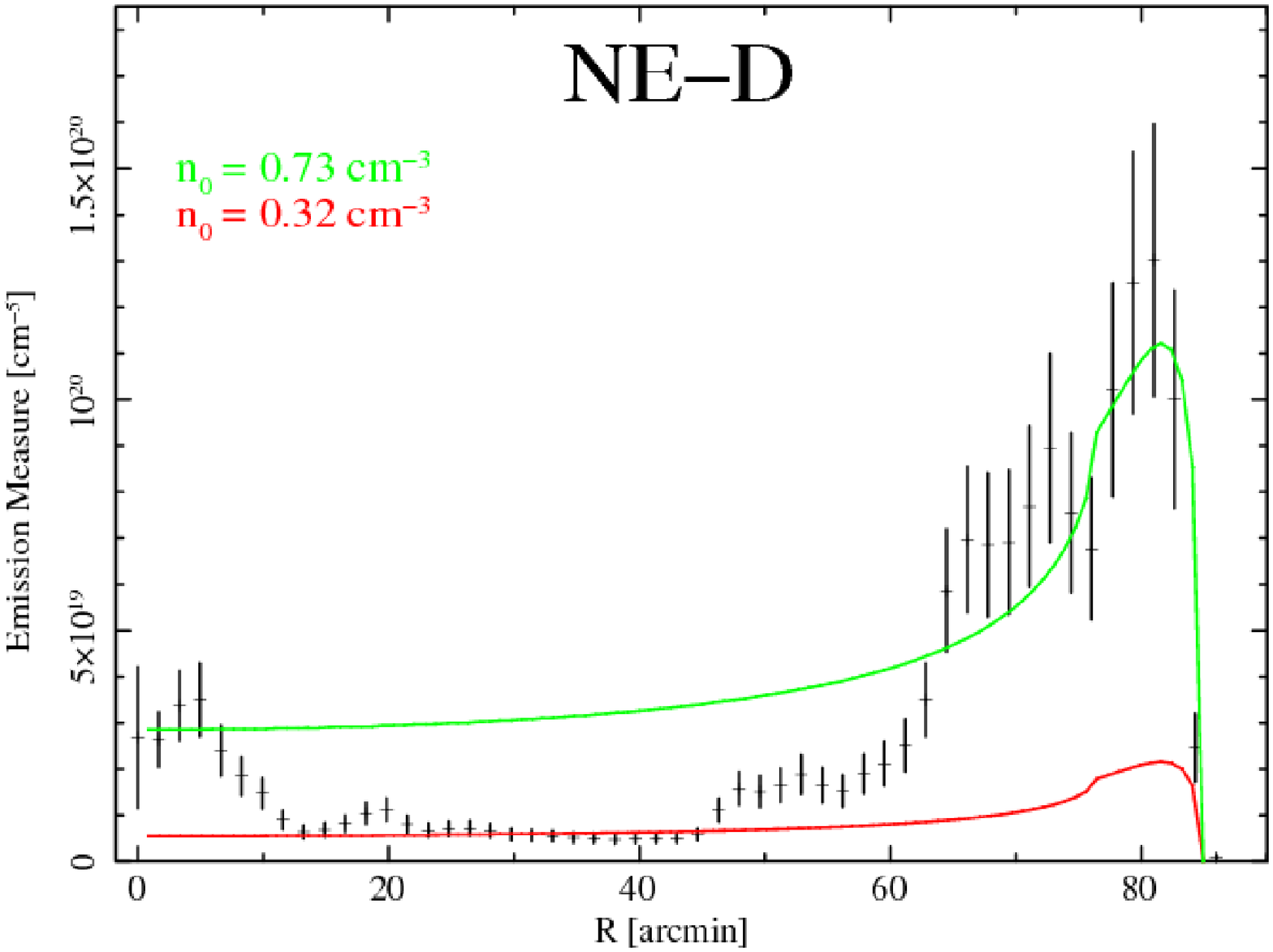}
    \includegraphics[width=75mm]{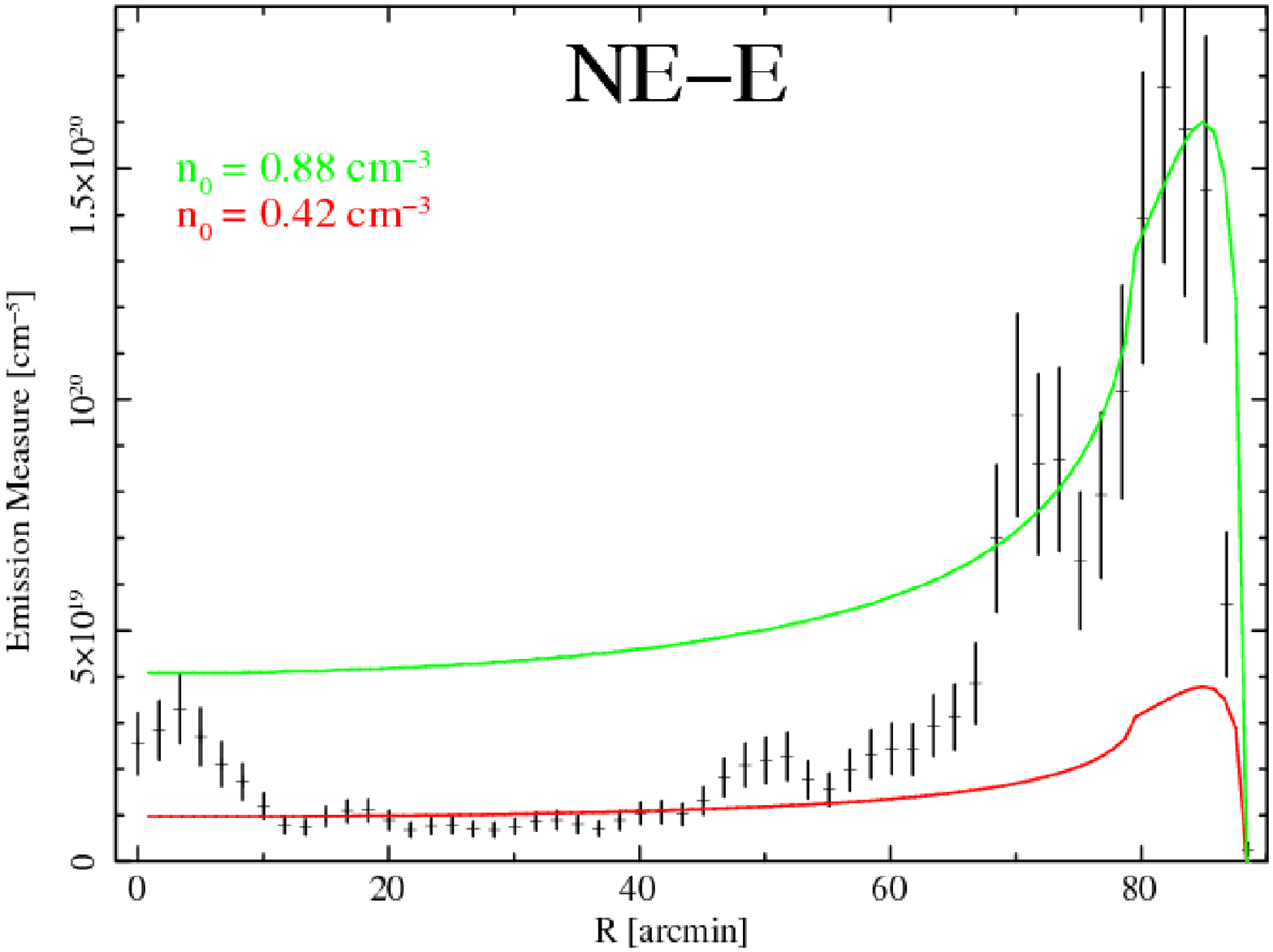}
    \includegraphics[width=75mm]{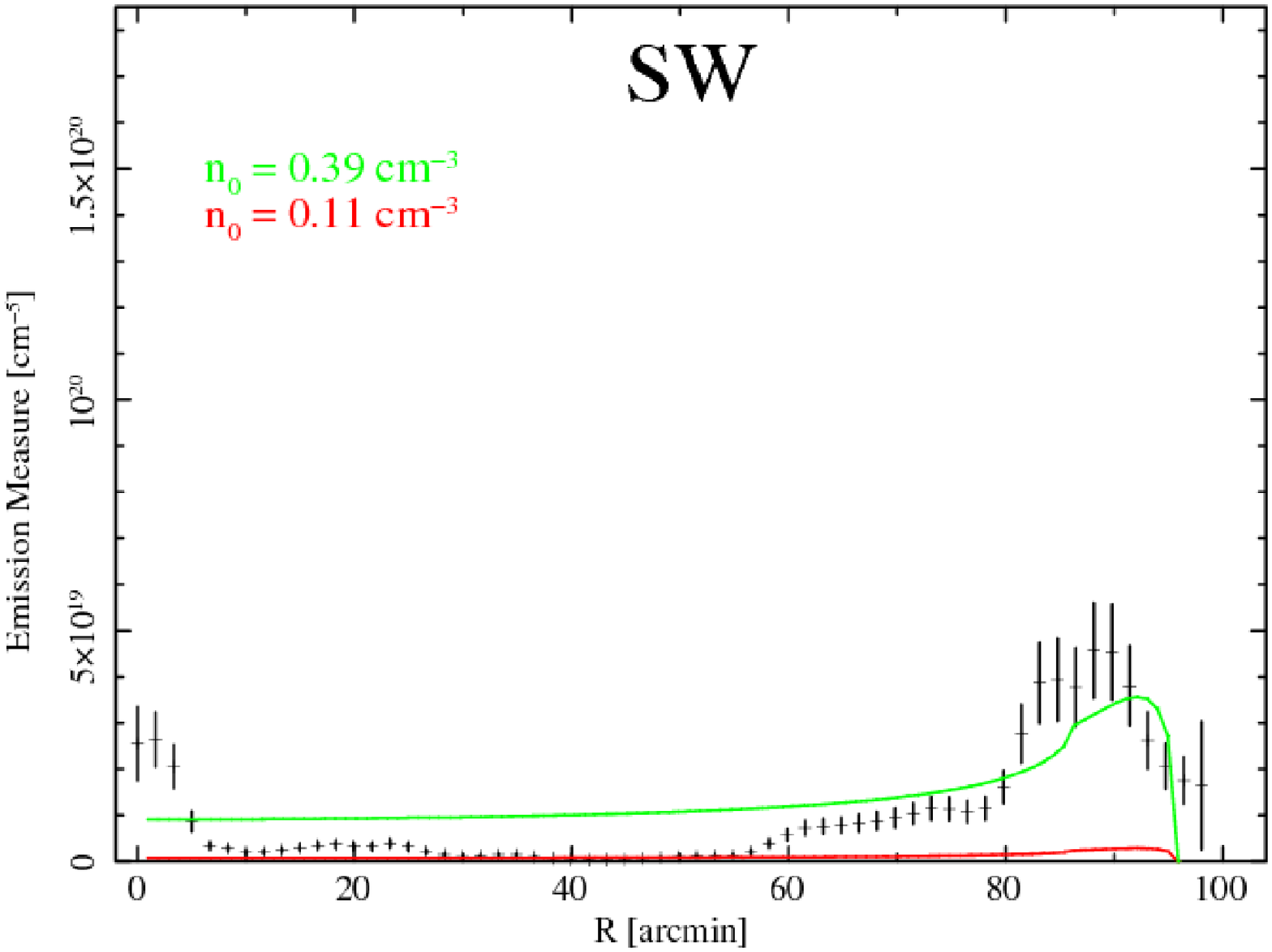}
  \end{center}
  \caption{EM profiles as a function of $R$ calculated from the data in the rectangular regions shown in figure \ref{fig:EM_region}. The EM profiles based on the Sedov model and the estimated ambient densities $n_0$ are shown in red and green (see text).}\label{fig:EM_plots}
\end{figure}

In order to estimate the ambient density more quantitatively, we calculated the EM of the low-$kT_e$ component and plotted it as a function of $R$.
Figure \ref{fig:EM_region} shows the EM distribution of the low-$kT_e$ component.
We plotted the EM profiles from six rectangular regions with different azimuthal angles as shown in figure \ref{fig:EM_region} (NE-A to NE-E and SW).
That are shown in figure \ref{fig:EM_plots}.
We simulated the EM profile of the shell component derived from the Sedov solution with different ambient density $n_0$ and estimated $n_0$ by comparing our observations with the EM models.
In this model, we assume the shock radius of the \object{Cygnus Loop} to be 13 pc and the ejecta is filled in 90$\%$ of it.
The results are shown in figure \ref{fig:EM_plots} with red lines.
We also show the best-fit models using the data only in the limb-brightening regions with green lines.
As for the northeast regions, the EM profiles inside the Loop are close to the models of $n_0$=0.3-0.4 cm$^{-3}$ (red) while the EM values at the limb-brightening regions are higher than these models.
On the other hand, applying the data only in the limb-brightening regions (green), $n_0$ increases to 0.7-0.9 cm$^{-3}$.
In any case, there are no Sedov models which agree with the EM profiles of the northeast part of the \object{Cygnus Loop}.
The result is the same as the case of the southwest region while the ambient density $n_0$ is less than half of the northeast results.
These results clearly show the \object{Cygnus Loop} can not be explained by a simple Sedov evolution model.
\citet{McCray79} proposed that the \object{Cygnus Loop}'s SN explosion had occurred in a preexisting cavity and some other studies also supported it \citep{Hester86, Hester94, Levenson97}.  
Considering their results, it is natural that the EM distribution disagrees with a simple Sedov model, and thus, we concluded that our result also supports the cavity explosion as the origin of the \object{Cygnus Loop} from the standpoint of the X-ray spectral analysis.
It should be noted that the Cygnus Loop is almost perfect circular in shape, although the EM (or flux) is globally higher in the northeast than that in the southwest. This fact strongly suggests that the northeast and the southwest blast waves should have hit the cavity wall very recently, and that the cavity-wall density is higher in the northeast than that in the southwest.

\section{Conclusion}

By analyzing the X-ray spectra, we clearly distinguished the ISM component from the ejecta component, and established a method to investigate the line-of-sight shell structure.
From the flux distribution of the ISM component, we found three low-flux regions in the FOV;
one is a well-known southwest blowout which is evidence of the cavity-wall break, and we also found other low-flux regions  at the west and the northeast of the \object{Cygnus Loop} center.
From the EM distribution of the ISM component, we support that the \object{Cygnus Loop} is originated from a cavity explosion.
Then, the ISM component, or cavity wall does not have an uniform structure but has a lot of breaks or tenuous regions.
We also found that the condition of the surrounding cavity wall is not uniform;  the density of it is globally higher in the northeast than that in the southwest.

\acknowledgments
H.U. would like to thank Professor Jacco Vink and his students for many useful discussions and their hospitality at Utrecht University.
This work is partly supported by a Grant-in-Aid for Scientific Research by the Ministry of Education, Culture, Sports, Science and Technology (16002004).  
H.U. and S.K. are supported by JSPS Research Fellowship for Young Scientists. 

\clearpage

\clearpage

\clearpage

\begin{deluxetable}{lcccc}
\tablewidth{0pt}
\tablecaption{Summary of the 41 observations\label{tab:sum}}
\tablehead{\colhead{Obs. ID} & \colhead{Obs. Date}& \colhead{RA, DEC (J2000)} & \colhead{Position Angle} & \colhead{Effective Exposure}}
\startdata
\sidehead{\textit{Suzaku Observations}}

501012010  (P1) & 2007-11-14 &  20$^{\mathrm h}$54$^{\mathrm m}$07.6$^{\mathrm s}$, 31\arcdeg57\arcmin22.0\arcsec & 240$^\circ$.0 & 9.8 ksec\\

501013010  (P2) & 2007-11-14 &  20$^{\mathrm h}$53$^{\mathrm m}$08.5$^{\mathrm s}$, 31\arcdeg45\arcmin40.3\arcsec & 240$^\circ$.0 & 16.4 ksec\\

501014010  (P3) & 2007-11-14 &  20$^{\mathrm h}$52$^{\mathrm m}$09.9$^{\mathrm s}$, 31\arcdeg36\arcmin43.4\arcsec & 240$^\circ$.0 & 16.9 ksec\\

501015010  (P4) & 2007-11-14 &  20$^{\mathrm h}$51$^{\mathrm m}$11.8$^{\mathrm s}$, 31\arcdeg22\arcmin08.4\arcsec & 240$^\circ$.0 & 18.3 ksec\\

501016010  (P5) & 2007-11-15 &  20$^{\mathrm h}$50$^{\mathrm m}$11.3$^{\mathrm s}$, 31\arcdeg10\arcmin48.0\arcsec & 240$^\circ$.0 & 19.3 ksec\\

501017010  (P6) & 2007-11-11 &  20$^{\mathrm h}$49$^{\mathrm m}$11.3$^{\mathrm s}$, 30\arcdeg59\arcmin27.6\arcsec & 240$^\circ$.0 & 28.7 ksec\\

501018010  (P7) & 2007-11-12 &  20$^{\mathrm h}$48$^{\mathrm m}$18.7$^{\mathrm s}$, 30\arcdeg46\arcmin33.6\arcsec & 240$^\circ$.0 & 21.0 ksec\\

501028010  (P8) & 2006-05-13 &  20$^{\mathrm h}$55$^{\mathrm m}$56.3$^{\mathrm s}$, 31\arcdeg28\arcmin56.2\arcsec & 62$^\circ$.5 & 4.9 ksec\\

501019010  (P9) & 2007-11-12 &  20$^{\mathrm h}$47$^{\mathrm m}$14.2$^{\mathrm s}$, 30\arcdeg36\arcmin10.8\arcsec & 240$^\circ$.0 & 16.2 ksec\\

501020010  (P10) & 2007-11-13 & 20$^{\mathrm h}$46$^{\mathrm m}$20.8$^{\mathrm s}$, 30\arcdeg23\arcmin22.6\arcsec & 240$^\circ$.0 & 14.7 ksec\\

503055010  (P11) & 2008-05-09 &  20$^{\mathrm h}$49$^{\mathrm m}$48.7$^{\mathrm s}$, 31\arcdeg30\arcmin18.0\arcsec & 50$^\circ$.0 & 22.2 ksec\\

501029010  (P12) & 2006-05-09 &  20$^{\mathrm h}$55$^{\mathrm m}$00.0$^{\mathrm s}$, 31\arcdeg15\arcmin46.8\arcsec & 62$^\circ$.1 & 13.2 ksec\\

501030010  (P13) & 2006-05-10 &  20$^{\mathrm h}$53$^{\mathrm m}$59.3$^{\mathrm s}$, 31\arcdeg03\arcmin39.6\arcsec & 68$^\circ$.2 & 13.9 ksec\\

501031010  (P14) & 2006-05-12 &  20$^{\mathrm h}$52$^{\mathrm m}$58.8$^{\mathrm s}$, 30\arcdeg51\arcmin32.4\arcsec & 62$^\circ$.4 & 18.2 ksec\\

501032010  (P15) & 2006-05-25 &  20$^{\mathrm h}$51$^{\mathrm m}$58.6$^{\mathrm s}$, 30\arcdeg39\arcmin10.8\arcsec & 62$^\circ$.0 & 17.4 ksec\\

501033010  (P16) & 2006-05-22 &  20$^{\mathrm h}$50$^{\mathrm m}$58.8$^{\mathrm s}$, 30\arcdeg27\arcmin00.0\arcsec & 62$^\circ$.0 & 20.0 ksec\\

501034010  (P17) & 2006-05-22 &  20$^{\mathrm h}$48$^{\mathrm m}$49.7$^{\mathrm s}$, 30\arcdeg00\arcmin21.6\arcsec & 62$^\circ$.0 & 13.9 ksec\\

501035010  (P18) & 2006-12-18 &  20$^{\mathrm h}$48$^{\mathrm m}$16.2$^{\mathrm s}$, 29\arcdeg42\arcmin07.2\arcsec & 237$^\circ$.5 & 11.2 ksec\\

501036010  (P19) & 2006-12-18 &  20$^{\mathrm h}$47$^{\mathrm m}$17.3$^{\mathrm s}$, 30\arcdeg04\arcmin21.4\arcsec & 237$^\circ$.5 & 11.8 ksec\\

503056010  (P20) & 2008-05-10 &  20$^{\mathrm h}$48$^{\mathrm m}$00.0$^{\mathrm s}$, 31\arcdeg10\arcmin30.0\arcsec & 50$^\circ$.0 & 22.5 ksec\\

503057010  (P21) & 2008-06-02 &  20$^{\mathrm h}$52$^{\mathrm m}$43.8$^{\mathrm s}$, 32\arcdeg26\arcmin19.0\arcsec & 61$^\circ$.9 & 16.2 ksec\\

503058010  (P22) & 2008-06-03 &  20$^{\mathrm h}$51$^{\mathrm m}$17.2$^{\mathrm s}$, 32\arcdeg25\arcmin24.6\arcsec & 61$^\circ$.4 & 19.3 ksec\\

503059010  (P23) & 2008-06-03 &  20$^{\mathrm h}$49$^{\mathrm m}$50.6$^{\mathrm s}$, 32\arcdeg21\arcmin50.8\arcsec & 61$^\circ$.9 & 19.5 ksec\\

503060010  (P24) & 2008-06-04 &  20$^{\mathrm h}$48$^{\mathrm m}$28.2$^{\mathrm s}$, 32\arcdeg17\arcmin44.5\arcsec & 61$^\circ$.4 & 18.5 ksec\\

503061010  (P25) & 2008-06-04 &  20$^{\mathrm h}$47$^{\mathrm m}$22.7$^{\mathrm s}$, 32\arcdeg10\arcmin22.8\arcsec & 60$^\circ$.9 & 26.0 ksec\\

503062010  (P26) & 2008-05-13 &  20$^{\mathrm h}$56$^{\mathrm m}$26.5$^{\mathrm s}$, 30\arcdeg19\arcmin55.2\arcsec & 49$^\circ$.8 & 16.9 ksec\\

503063010  (P27) & 2008-05-13 &  20$^{\mathrm h}$55$^{\mathrm m}$16.3$^{\mathrm s}$, 30\arcdeg01\arcmin44.0\arcsec & 49$^\circ$.6 & 22.8 ksec\\

503064010  (P28) & 2008-05-14 &  20$^{\mathrm h}$53$^{\mathrm m}$51.6$^{\mathrm s}$, 29\arcdeg54\arcmin42.5\arcsec & 49$^\circ$.1 & 18.2 ksec\\

500020010  (NE1) & 2005-11-23 &  20$^{\mathrm h}$56$^{\mathrm m}$48.9$^{\mathrm s}$, 31\arcdeg56\arcmin54.8\arcsec & 223$^\circ$.0 & 20.4 ksec\\

500021010  (NE2) & 2005-11-24 &  20$^{\mathrm h}$55$^{\mathrm m}$56.0$^{\mathrm s}$, 31\arcdeg56\arcmin53.2\arcsec & 223$^\circ$.0 & 21.4 ksec\\

500022010  (NE3) & 2005-11-29 &  20$^{\mathrm h}$55$^{\mathrm m}$05.6$^{\mathrm s}$, 32\arcdeg10\arcmin35.4\arcsec & 222$^\circ$.9 & 21.7 ksec\\

500023010  (NE4) & 2005-11-30 &  20$^{\mathrm h}$54$^{\mathrm m}$03.8$^{\mathrm s}$, 32\arcdeg21\arcmin47.9\arcsec & 221$^\circ$.2 & 25.3 ksec\\

\sidehead{\textit{XMM-Newton Observations}}

0082540101 (Pos-1) & 2002-11-25 & 20$^{\mathrm h}$55$^{\mathrm m}$23.6$^{\mathrm s}$, 31\arcdeg46\arcmin17.0\arcsec & 241$^\circ$.7 & 14.7 ksec\\

0082540201 (Pos-2) & 2002-12-03 & 20$^{\mathrm h}$54$^{\mathrm m}$07.2$^{\mathrm s}$, 31\arcdeg30\arcmin51.1\arcsec & 241$^\circ$.7 & 14.4 ksec\\

0082540301 (Pos-3) & 2002-12-05 & 20$^{\mathrm h}$52$^{\mathrm m}$51.1$^{\mathrm s}$, 31\arcdeg15\arcmin25.7\arcsec & 241$^\circ$.7 & 11.6 ksec\\

0082540401 (Pos-4) & 2002-12-07 & 20$^{\mathrm h}$51$^{\mathrm m}$34.7$^{\mathrm s}$, 31\arcdeg00\arcmin00.0\arcsec & 241$^\circ$.7 & 4.9 ksec\\

0082540501 (Pos-5) & 2002-12-09 & 20$^{\mathrm h}$50$^{\mathrm m}$18.4$^{\mathrm s}$, 30\arcdeg44\arcmin34.3\arcsec & 231$^\circ$.4 & 12.6 ksec\\

0082540601 (Pos-6) & 2002-12-11 & 20$^{\mathrm h}$49$^{\mathrm m}$02.0$^{\mathrm s}$, 30\arcdeg29\arcmin08.6\arcsec & 241$^\circ$.7 & 11.5 ksec\\

0082540701 (Pos-7) & 2002-12-13 & 20$^{\mathrm h}$47$^{\mathrm m}$45.8$^{\mathrm s}$, 30\arcdeg13\arcmin42.9\arcsec & 241$^\circ$.7 & 13.7 ksec\\

0405490101 (Pos-8) & 2006-05-13 & 20$^{\mathrm h}$50$^{\mathrm m}$32.2$^{\mathrm s}$, 30\arcdeg11\arcmin00.0\arcsec & 69$^\circ$.9 & 6.5 ksec\\

0405490201 (Pos-9) & 2006-05-13 & 20$^{\mathrm h}$49$^{\mathrm m}$54.2$^{\mathrm s}$, 29\arcdeg42\arcmin25.0\arcsec & 69$^\circ$.8 & 3.6 ksec\\

\enddata
\end{deluxetable}

\begin{table}
 \caption{Spectral fit parameters}\label{tab:spec}
  \begin{center}
    \begin{tabular}{ccccc}
       \tableline 
      \tableline
           & \multicolumn{2}{c}{\textit{single-component VNEI model}} & \multicolumn{2}{c}{\textit{two-component VNEI model}} \\
      \tableline
  & region A  & region B & region A  & region B\\
      \tableline
      N$\rm _H$ [10$^{20}$cm$^{-2}$] & 1.8 $\pm$ 0.3 & 3.4 $\pm$ 0.3 & 5.2 $\pm$ 0.2 & 7.0 $\pm$ 0.3 \\
  &   & & \multicolumn{2}{c}{\textit{Low-$kT_e$ component:}} \\
      \ \ $kT_e$ [keV] & 0.59 $\pm$ 0.03 & 0.42 $\pm$ 0.02 & 0.24 $\pm$ 0.01 & 0.12 $\pm$ 0.01 \\ 
      \ \ C &  0.24 $\pm$ 0.05 & 0.96 $\pm$ 0.21 & \multicolumn{2}{c}{0.27 (fixed)}\\
      \ \ N &  0.22 $\pm$ 0.05 & 0.09 $\pm$ 0.03 & \multicolumn{2}{c}{0.10 (fixed)}\\
      \ \ O &  0.23 $\pm$ 0.02 & 0.14 $\pm$ 0.02 & \multicolumn{2}{c}{0.11 (fixed)}\\
      \ \ Ne & 0.44 $\pm$ 0.04 & 0.31 $\pm$ 0.03 &  \multicolumn{2}{c}{0.21 (fixed)}\\
      \ \ Mg & 0.26 $\pm$ 0.03 & 0.24 $\pm$ 0.03 &  \multicolumn{2}{c}{0.17 (fixed)}\\
      \ \ Si & 0.27 $\pm$ 0.06 & 0.30 $\pm$ 0.06 &   \multicolumn{2}{c}{0.34 (fixed)}\\
      \ \ S &  (=Si) & (=Si) & \multicolumn{2}{c}{0.17 (fixed)}\\
      \ \ Fe(=Ni) &  0.36 $\pm$ 0.04 & 0.22 $\pm$ 0.02 & \multicolumn{2}{c}{0.20 (fixed)}\\
      \ \ log $\tau$ & 10.42 $\pm$ 0.03 & 10.82 $^{+ 0.07}_{- 0.08}$ & 11.32 $^{+ 0.12}_{- 0.16}$  & $<$ 12\\
\ \ flux [counts cm$^{-2}$s$^{-1}$arcmin$^{-2}$] & $8.90 \times 10^{-4}$ & $4.34 \times 10^{-4}$ & $7.49 \times 10^{-4}$ & $2.18 \times 10^{-4}$\\
  &   & & \multicolumn{2}{c}{\textit{High-$kT_e$ component:}} \\
      \ \ $kT_e$ [keV] & \nodata & \nodata & 0.88 $\pm$ 0.13  & 0.43 $\pm$ 0.02\\ 
      \ \ O(=C=N) & \nodata & \nodata  & 0.34 $\pm$ 0.13 & 0.38 $\pm$ 0.07\\
      \ \ Ne & \nodata & \nodata  & 0.82 $\pm$ 0.26  & 0.74 $\pm$ 0.12\\
      \ \ Mg & \nodata & \nodata  & 0.56 $\pm$ 0.19 & 0.55 $\pm$ 0.10\\
      \ \ Si(=S) & \nodata & \nodata  & 1.28 $\pm$ 0.42 & 0.79 $\pm$ 0.16\\
      \ \ Fe(=Ni) & \nodata & \nodata  & $<$ 1.43 & 0.48 $\pm$ 0.08\\
      \ \ log $\tau$ & \nodata & \nodata  & 10.66  $^{+ 0.09}_{- 0.12}$ & 11.11 $\pm$ 0.06\\
\ \ flux [counts cm$^{-2}$s$^{-1}$arcmin$^{-2}$] & \nodata & \nodata  & $1.41 \times 10^{-4}$ & $2.16 \times 10^{-4}$\\
$\chi ^2$/dof & 1043/739 & 728/548 & 868/738 & 637/547\\
      \tableline
    \end{tabular}
\tablecomments{Other elements are fixed to solar values.}
 \end{center}
\end{table}

\clearpage

\end{document}